\documentclass[conference]{IEEEtran}
\usepackage[english]{babel}
\usepackage[export]{adjustbox}
\usepackage[justification=centering]{caption}
\usepackage[section]{placeins}
\usepackage[T1]{fontenc}
\usepackage[table]{xcolor}
\usepackage[utf8]{inputenc}
\usepackage{algorithm,algcompatible,amsmath}
\usepackage{amsfonts}
\usepackage{amssymb}
\usepackage{balance}
\usepackage{blindtext}
\usepackage{cases}
\usepackage{cite}
\usepackage{epsfig}
\usepackage{epstopdf}
\usepackage{fancyhdr}
\usepackage{float}
\usepackage{graphicx}
\usepackage{latexsym}
\usepackage{lettrine}
\usepackage{lipsum, color}
\usepackage{makecell}
\usepackage{mathtools, cuted}
\usepackage{multicol,lipsum}
\usepackage{multirow}
\usepackage{pdfpages}
\usepackage{pifont}
\usepackage{pst-all}
\usepackage{pstricks}
\usepackage{resizegather}
\usepackage{setspace}
\usepackage{stfloats}
\usepackage{subcaption}
\usepackage{upgreek}

\algnewcommand\INPUT{\item[\textbf{Input:}]}%
\algnewcommand\OUTPUT{\item[\textbf{Output:}]}%
\algnewcommand\FIND{\item[\textbf{Find}]}
\algnewcommand\INT{\item[\textbf{Initialize}]}
\algnewcommand\REE{\item[\textbf{Repeat}]}
\algnewcommand\IFF{\item[\textbf{If}]}
\algnewcommand\SD{\item[\textbf{For}]}

\allowdisplaybreaks

\graphicspath{{myfig/}}

\makeatletter

\IEEEoverridecommandlockouts
\vspace{-0.2in}
\title{Sum Rate Fairness Trade-off-based Resource Allocation Technique for  MISO NOMA Systems}
\vspace{-0.1in} 
\author{\IEEEauthorblockN{Haitham Al-Obiedollah\IEEEauthorrefmark{1}, Kanapathippillai Cumanan\IEEEauthorrefmark{1}, Jeyarajan Thiyagalingam\IEEEauthorrefmark{2},  Alister G. Burr\IEEEauthorrefmark{1},\\ Zhiguo Ding\IEEEauthorrefmark{3}, and Octavia A. Dobre\IEEEauthorrefmark{4}}\\
\IEEEauthorblockA{\IEEEauthorrefmark{1}Department of Electronic Engineering, University of York, York, YO10 5DD, UK\\
\IEEEauthorblockA{\IEEEauthorrefmark{2}STFC, Rutherford Appleton Laboratory, Oxford, OX11 0QX, UK
\\
\IEEEauthorblockA{\IEEEauthorrefmark{3} School of Electrical and Electronic Engineering, The University of Manchester,
Manchester, UK}\vspace{-0.2in} \\
\IEEEauthorblockA{\IEEEauthorrefmark{4}Department of Electrical and Computer Engineering, Memorial University, St. John’s,  Canada}\vspace{-0.2in} \\
Email: \{hma534, kanapathippillai.cumanan, alister.burr\}@york.ac.uk\IEEEauthorrefmark{1},\\  t.jeyan@stfc.ac.uk\IEEEauthorrefmark{2},  zhiguo.ding@manchester.ac.uk\IEEEauthorrefmark{3}, odobre@mun.ca\IEEEauthorrefmark{4}}
}}

\begin{document}
 
\maketitle
 
\linespread{.91}
\begin{abstract}
 In this paper, we propose a beamforming design  that  jointly considers two conflicting performance metrics, namely  the sum rate  and  fairness,   for a multiple-input single-output    non-orthogonal multiple access system.  Unlike the conventional rate-aware beamforming designs, the proposed approach  has the flexibility to   assign different weights to the objectives (i.e., sum rate and fairness) according to the network requirements and the channel conditions. In particular, the proposed design is first formulated as a multi-objective optimization  problem, and  subsequently mapped to a single objective optimization (SOO) problem by exploiting the weighted sum approach combined with a prior articulation method. As the resulting SOO problem is non-convex, we use the sequential convex approximation  technique, which introduces multiple slack variables, to solve the overall problem. Simulation results are provided to demonstrate the  performance and the effectiveness of the proposed approach along with detailed comparisons with conventional rate-aware-based beamforming designs.
\end{abstract}
\begin{IEEEkeywords} Beamforming design,  multi-objective optimization, non-orthogonal multiple access,  Pareto-optimal.
\end{IEEEkeywords}
\IEEEpeerreviewmaketitle
\graphicspath{{myfig/}}

\section{Introduction}

Non-orthogonal multiple access (NOMA) has been proposed as a novel multiple access scheme to overcome the relatively poor spectral-efficiency of   the   conventional orthogonal multiple access (OMA) schemes \cite{fnoma}, \cite{noma4}.  In the power-domain NOMA,      superposition coding (SC) is employed   to encode different signals   with  different power levels through power domain multiplexing \cite{noma4}. In particular,  the users with lower channel   gains are assigned with higher power levels compared to those with higher channel gains \cite{islam2018}. At the receiver end,  
 stronger users exploit successive interference cancellation (SIC) to subtract
  the interference from weaker users  before detecting their own signals \cite{noma4}.  This multiple access technique, along with other  disruptive technologies, such as massive multiple-input multiple-output (MIMO) and mm Wave communication, has the potential to  further improve the performance of the  fifth generation (5G) and beyond wireless networks \cite{mimonoma} \cite{misonoma}.  Recently, different rate-aware beamforming designs   have been proposed for multiple-input single-output   (MISO) NOMA systems. For example, the sum rate maximization (SRM)-based design  maximizes the sum rate of all users in the cell, however,   without taking the  individual users rates  into account~\cite{hanif}. This approach significantly  degrades the rates of the users with weaker channel conditions. To overcome this issue, in  \cite{wsrmnoma},  a rate-fairness-based design has been developed through the weighted sum-rate maximization (WSRM). In WSRM, higher weights are assigned to weaker users' rates to maintain the fairness between users in terms of their achievable   rates. 
However, none of  these conventional rate-aware-based designs   consider  either the instantaneous rate-requirements of the users, or the     variations of the users'  channel strengths due to the mobility of the users.  For example, SRM-based design is an appropriate beamforming design when the users have similar channel strengths. However, the SRM-based design is not capable of achieving a reasonable throughput for all users in a system where the channel strengths of the users vary significantly. Such cases, the weakest user will suffer from low quality of service.    In particular, both performance metrics, the sum rate and the fairness among users are crucial performance metrics that have to be considered in 5G and beyond wireless networks   \cite{cuma9} \cite{cuma11} \cite{cuma12}. Hence, the base station (BS) should have the flexibility to intelligently decide whether it needs to     maximize the sum rate   or  the fairness among users, or to strike a good balance between  them.
 The fairness index (FI) has been used to measure the  fairness  between users in terms of their achievable rates \cite{fairindex}. In particular, the FI of the system with $K$ users is defined as follows \cite{fairness}, \cite{cuma7}:
\begin{equation}\label{fairness index}
 \text{FI} =\frac{(\sum_{i=1}^{K} R_i)^2}{K \sum_{i=1}^{K} R_i ^2},
\end{equation}
where $R_{i}$ denotes the achieved rate of the $i^{th}$ user $(u_{i})$. The best fairness can be achieved  when  FI is one.   Note that the    FI and  sum rate are conflicting performance metrics, which means that maximizing the sum rate will degrade the FI, and vice versa, especially with users with significantly different channel strengths.\\
\indent Motivated by this discussion,  we propose a novel  beamforming design   that jointly considers the   conflicting performance metrics, i.e.,   the  sum rate and  fairness in a  MISO NOMA system. In this joint design, the  BS   decides the importance of each performance metric (i.e., sum rate and FI) through assigning  a weight factor  for each objective  based on   the service requirements and the channel conditions of the users. For instance, the BS will consider the fairness with higher weight  when the channel strengths of the users are significantly different and the users expect to achieve the same quality of service in terms of their achievable throughput.   On the other hand,   more weight will be assigned  for   the sum rate when the users have  similar channel conditions. Furthermore, this joint design has the capability to strike a balance between the sum rate and  fairness through assigning appropriate weights to each performance metric. 
\indent In particular, we formulate this trade-off-based design as  a  multi-objective optimization (MOO) problem, which  is difficult to solve for an optimal solution. Therefore, we rewrite this MOO problem as a single objective optimization (SOO) problem  by employing a prior articulation method combined with the weighted sum approach\cite{MO3}-\!\cite{MO2}. In the prior articulation scheme,  the  BS  decides the  weight  of each objective in the MOO problem prior to designing
 the beamforming vectors.  A weighted sum single objective function is used to represent the multi-objective functions  \cite{MO3}. However,  the obtained SOO is non-convex and we employ sequential convex approximations (SCA) to solve it.\\
\indent The remainder of the paper is organized as follows.  Section \ref{sec2}  presents  the system model and   the problem formulation. Section \ref{sec3} demonstrates  the proposed technique to solve the developed optimization problem.  Section \ref{sec4}  provides simulation results to validate the effectiveness of the proposed beamforming design  by comparing its performance with different beamforming techniques available in the literature. Finally,   Section \ref{sec5}   concludes the paper.

\subsection*{Notations}
We use lower case boldface letters for vectors and upper case  boldface letters for matrices. $(\cdot )^H $   denotes complex conjugate transpose.    $\Re(\cdot)$ and $\Im(\cdot)$  stand for  real and imaginary parts of a complex number, respectively. The symbols $\mathbb{C}^{N}$ and $\mathbb{R}^{N}$ denote  $N$-dimensional  complex and real spaces, respectively.      $||\cdot||_2 $  and $| \cdot| $ represent  the Euclidean norm of  a vector and the absolute value of a complex number, respectively. $\mathbf{x} \succ0$ means that all the elements in the  vector  $\mathbf{x}$ are greater than zero.
\section{System Model And Problem Formulation}\label{sec2}
\subsection{System Model}
 
In this paper, we consider  the downlink transmission of a  MISO  NOMA system,  in which a  BS    equipped with  $N$  antennas simultaneously   transmits signals to $K$ single-antenna users. The  transmitted signal from the BS can be written as 
\begin{equation}\label{super}
\mathbf{x}=\sum_{i=1}^{K}\mathbf{w}_is_i,
\end{equation}
where $s_{i}$ and $\mathbf{w}_i$ $ \in$ $\mathbb{C}^{N\times1}$ denote the signal intended for the $i^{th}$ user $u_{i}$ and the corresponding beamforming vector, respectively. The received signal at $u_i$   can be written as 
\begin{equation}
y_i=  {\mathbf{h}_i}^H \mathbf{w}_i s_i + \sum_{j=1, j\neq i}^{K}{\mathbf{h}_i}^H \mathbf{w}_j s_j +n_i, 
\end{equation}
where $\mathbf{h}_i$  $ \in$ $\mathbb{C}^{N\times1}, \forall i \in \mathcal{K}\overset{\bigtriangleup}{=}\{1,\cdots,K\}$ represents the channel coefficient vector between $u_i$   and the BS. These channel coefficients are modelled as   $ \mathbf{h}_i= \sqrt{d_i^{-\kappa}}\mathbf{g}_i$, where $ \kappa , d_{i}$ and $ g_{i} $  denote the path loss exponent, the distance between $u_i$  and the  BS in meters, and the small scale fading, respectively. In addition, $n_i\sim \mathcal{CN}(0,\sigma_i^2)$ represents the additive white Gaussian noise with zero-mean and variance  $\sigma_i^2$.  In  power-domain NOMA,  user  ordering plays a crucial role for the performance of  NOMA \cite{cuma8},   which  can only be determined through exhaustive search to achieve the optimal performance.  However, for the sake of simplicity, we order the users based on their channel strengths   as   follows:
\begin{equation}\label{order}
||\mathbf{h}_1||_2^2\leq ||\mathbf{h}_2||_2^2\leq\cdots\leq||\mathbf{h}_K||_2^2, 
\end{equation} 
  where $u_1$ and $u_{K}$ are refer to the weakest and strongest users in the system.
Based on this user ordering in (\ref{order}),  $u_i$ has the capability to perform    SIC   through   decoding and subtracting
 the signals intended for the $u_1,u_2,\cdots,u_{i-1}$   users     prior to decoding its own signal\cite{misonoma}. Therefore, the received signal at  $u_i$    after eliminating the first $i-1$ users' signals using  SIC  can be written as \cite{wcnc1}
\begin{equation} 
\overset{\sim}{y_i}=  \underbrace{{\mathbf{h}_i}^H \mathbf{w}_i s_i}_{\text{intended signal }} +\underbrace{\sum_{j=i+1}^{K} \mathbf{h}_i^H \mathbf{w}_j s_j }_{ \text{ interference }  }+  \underbrace{n_i}_{\text{ noise }}, \forall i \in \mathcal{K}.
\end{equation}
In particular, the signal intended for $u_i $  will be decoded at $u_k$   (i.e., $k\geq i$ ) with the following  signal to interference plus  noise   ratio  ${SINR}_i^{(k)}$:
\begin{equation}
{SINR}_i^{(k)}=\frac{|\mathbf{h}_k^H \mathbf{w}_i|^2}{\sum_{j=i+1}^{K}|\mathbf{h}_k^H \mathbf{w}_j|^2 +\sigma_k^2}, \quad \forall i \in \mathcal{K}, k\geq i.
\end{equation}
In order to decode  the $i^{th}$   user signal at different users, the $SINR$ of that signal should be more than a  certain threshold. This imposes a condition   ${SINR}_i^{(k)} (k \geq i)$ at all the strong users that they should satisfy the predefined SINR threshold to achieve a particular rate. Therefore, the  ${SINR}$ of the signal  intended for $u_i$ can be defined as \cite{hanif}
\begin{equation}\label{sinr}
 {SINR}_i=\text{min} \{ {SINR}_i^{(i)}, \cdots, {SINR}_i^{(K)}\},  \forall i \in \mathcal{K},
 \end{equation}
and  the rate of $u_i$  can be defined as
\begin{equation}\label{rate}
\ {R}_i=B_w\log(1+ {SINR}_i),\quad \forall i \in \mathcal{K},
\end{equation}
where $B_w$ is the available bandwidth for transmission,  which is      assumed to be one. Furthermore,  the sum rate of this MISO NOMA system  is given by
\begin{equation}\label{sumrate}
 {R} =\sum_{i=1}^{K} {R}_i. 
\end{equation} 
In order to ensure that SIC is successfully implemented at all  strong users, and to assign more power levels to the  weaker users  based on NOMA,   the following
SIC constraints should be included \cite{fayzeh}: 
\begin{equation}\label{channels}
 \vert \mathbf{h}_i^H\mathbf{w}_1 \vert^2 \geq \cdots\geq\vert \mathbf{h}_i^H \mathbf{w}_{K}\vert^2,   \forall i \in \mathcal{K}.
\end{equation}
In addition, the  transmit power ($P_{tr}$)  should not exceed  the available power budget ($P_{ava}$) at the BS, which  can be mathematically  formulated as the following constraint:
\begin{equation}\label{power}
P_{tr}=\sum_{i=1}^{K}\vert \vert \mathbf{w}_i \vert \vert ^2_2 \leq P_{ava}.
\end{equation}

\subsection{Problem Formulation}
For the sake of notation simplicity, we denote the sum rate by $f_1(\{\mathbf{w}_i\}_{i=1}^{K})$  (i.e.,  $f_1(\{\mathbf{w}_i\}_{i=1}^{K})  = R$), whereas FI is represented by $f_2(\{\mathbf{w}_i\}_{i=1}^{K}) $ (i.e., $f_2(\{\mathbf{w}_i\}_{i=1}^{K}) $ = FI).   In this work, we aim to jointly maximize   the conflicting objectives  (i.e., maximize $f_1(\{\mathbf{w}_i\}_{i=1}^{K})$ and $f_2(\{\mathbf{w}_i\}_{i=1}^{K})$) subject to SIC  and total transmit power constraints. This could be mathematically formulated as the  following MOO problem:
\begin{subequations}\label{MOO}
\begin{align}
P_1:
 &\underset{\{\mathbf{w}_i\}_{i=1}^{K} }{\text{max }}\!\!\!\!\!\!\!\!\!
& & \! \! \! \! \! \! \! \! \! \! \mathbf{f}(\{\mathbf{w}_i  \}_{i=1}^{K})  \\
& ~~\text{s.t.}
& & \! \! \! \! \! \! \! \! \! \!(\ref{channels}),(\ref{power}),  
\end{align}
\end{subequations}  
where   $\mathbf{f}(\{\mathbf{w}_i\}_{i=1}^{K}) $  denotes the vector which  consists of  the both objective functions  (i.e., $\mathbf{f}(\{\mathbf{w}_i  \}_{i=1}^{K}) = [f_1(\{\mathbf{w}_i  \}_{i=1}^{K}) ~  f_2(\{\mathbf{w}_i  \}_{i=1}^{K})]^T$). In fact, there exists
 no  single global optimal solution that simultaneously  maximizes $f_1(\{\mathbf{w}_i\}_{i=1}^{K})$  and $f_2(\{\mathbf{w}_i\}_{i=1}^{K})$ together. Therefore, to handle
  such a problem, the designers search for the best trade-off solutions according to the network conditions, which are named as the Pareto optimal solutions (non-dominated solutions) \cite{MO2}. In particular, a feasible solution ${\{\mathbf{w}_i^*\}_{i=1}^{K}}$ is called  a Pareto optimal solution if and only if  there exists
   no other feasible solution  ${\{\mathbf{w}_i^{'}  \}_{i=1}^{K}}$ such that $\mathbf{f}(\{\mathbf{w}_i^{'}\}_{i=1}^{K}) \succ \mathbf{f}({\{\mathbf{w}_i^*\}_{i=1}^{K}}) $. The set of all Pareto-optimal solutions is called the Pareto front\cite{MO2}. 
Therefore, in the following section,  we develop an effective approach  to determine the Pareto-optimal solution of the MOO problem in (\ref{MOO}). 
\section{Proposed Methodology }\label{sec3}
In this section, we provide an effective  approach to solve the challenging MOO problem  $ P_1$. In particular,  this approach is developed by   combining both the objective functions in
$ P_1$   as a weighted single objective function,  which we  solve by using the SCA technique.
\subsection{Single Objective Optimization}
In order to  reformulate the optimization problem in  (\ref{MOO}) into a tractable SOO problem,  we   employ  the prior articulation scheme combining with the weighted sum approach. First, a weight factor ($\alpha_i$) is assigned to  each objective function  such that $\alpha_1+\alpha_2=1$, where $\alpha_i$ for ($i=1,2$)  reflects the importance of $f_i(\{\mathbf{w}_i\}_{i=1}^{K})$ in the overall MOO problem.
Next, both   objective functions in $\mathbf{f}(\{\mathbf{w}_i\}_{i=1}^{K})$ are combined into a  single objective function (utility function)\cite{MO3}. A number of  utility functions have been considered in the literature of MOO, however,  we  employ the weighted sum  approach here as it achieves the Pareto-optimal solutions\cite{MO2}. It is worth mentioning that we have to normalize each objective function by its maximum value (i.e.,   \textit{utopia point}) prior to adding them in order to get  an unit-less objective function with a maximum value of one for each normalized objective function. Based on this weighted sum-approach, the original MOO problem in (\ref{MOO}) can be reformulated as a SOO as follows:
\begin{subequations}\label{MOO2}
\begin{align}
 \overset{\sim}{P_1}:
 & \underset{\{\mathbf{w}_i\}_{i=1}^{K} }{\text{max }}
& &   \!\!\!\!\!\! \!\!   \alpha_1 f_1^*(\{\mathbf{w}_i\}_{i=1}^{K}) + \alpha_2  f_2^*(\{\mathbf{w}_i\}_{i=1}^{K})  \\
& \text{s.t.}
& &  \!\!\!\!\!\! \!\! (\ref{channels}), (\ref{power}),
\end{align}
\end{subequations}
where $f_i^*(\{\mathbf{w}_i\}_{i=1}^{K}) $ is the  normalized  objective function of $f_i(\{\mathbf{w}_i\}_{i=1}^{K}) $, which can be written  as  \cite{funtr}
\begin{equation}
f_i^*(\{\mathbf{w}_i\}_{i=1}^{K}) =\frac{f_i(\{\mathbf{w}_i\}_{i=1}^{K}) }{f_i^{max}},
\end{equation}
   where  $f_i^{max}$ represents the maximum value of $f_i(\{\mathbf{w}_i\}_{i=1}^{K}) $.  In particular,  the  maximum value of the FI is one (i.e., $f^{max}_2=1$). On the other hand, the sum rate function (i.e., $f_1(\{\mathbf{w}_i\}_{i=1}^{K}) $) should be normalized by its maximum value prior to solving the SOO $\overset{\sim}{P_1}$,  which can be determined by solving the following   SRM problem:
\begin{subequations}\label{Max}
\begin{align}
P_2:
&\underset{\{\mathbf{w}_i\}_{i=1}^{K} }{\text{max }}
& & \!\!\!\!\!\! \!\!\!\!\!\! \!\!\!\!\!\! \!\!\!\!\!\! \sum_{i=1}^{K} {R}_i \\
& ~~\text{s.t.}
& & \!\!\!\!\!\! \!\!\!\!\!\! \!\!\!\!\!\! \!\!\!\!\!\!(\ref{power}), (\ref{channels}).
\end{align}
\end{subequations}
The SRM in (\ref{Max}) can be solved  by exploiting the  minorization maximization algorithm\cite{hanif}. In the following subsection, we develop an effective approach to solve the SOO in (\ref{MOO2}).
It is worth
 to make two important observations regarding    the SOO  optimization problem $\overset{\sim}{P_1}$. Firstly, the weighted sum approach that is utilized to replace the original multi-objective functions by a single one produces the  Pareto-optimal  solutions of the original optimization problem $P_1$ \cite{MO2}. Secondly, the original MOO $P_1$ turns out to be the conventional SRM problem when $\alpha_1=1$. However, when $\alpha_2=1$,  $\overset{\sim}{P_1}$ becomes the max-min rate (MMR)  optimization problem. In particular, the MMR solution achieves the same rate (i.e., a unity FI)  for all the users in the system \cite{cuma}. However,  the WSRM   and proportional fairness (PF) problems could be formulated from the original optimization problem $\overset{\sim}{P_1}$ through appropriately scaling the weight factor between zero and one.

\subsection{Sequential Convex Approximation}
In this  subsection, we   solve the $ \overset{\sim}{P_1}$   by approximating the non-convex functions in the objective and the constraints as convex ones through using the SCA approach \cite{sqa}. To apply this approach, we introduce multiple slack variables ($\xi_1,\xi_2,\xi$) to represent the single objective function as follows:
\begin{subequations}
\begin{align}
\alpha_1 f_1^*(\{\mathbf{w}_i\}_{i=1}^{K}) \geq \xi_1,\\
 \alpha_2  f_2^*(\{\mathbf{w}_i\}_{i=1}^{K}) \geq \xi_2,\\
 \xi_1+\xi_2 \geq \xi.
\end{align}
\end{subequations}
Based on these new slack variables, the optimization problem in (\ref{MOO2}) can be rewritten as
\begin{subequations}\label{MOO3}
\begin{align}
\overset{\approx}{P_1}:
& \underset{\xi_1,\xi_2,\xi,\{{\mathbf{w}_i}\}_{i=1}^K}{\text{max}}
& & \!\!\!\!\!\! \!\!\!\!\!\! \!\!  \xi\\
& ~~~~~~~ \text{s.t.}
& &\!\!\!\!\!\! \!\!\!\!\!\! \!\! \xi_1+\xi_2 \geq \xi,\label{cons1}\\
&& &\!\!\!\!\!\! \!\!\!\!\!\! \!\! (1-\alpha) f_1^*(\{\mathbf{w}_i\}_{i=1}^{K})\geq \xi_1, \label{con1}\\
& & & \!\!\!\!\!\! \!\!\!\!\!\! \!\! \alpha f_2^*(\{\mathbf{w}_i\}_{i=1}^{K})\geq \xi_2,\label{con2}\\
& && \!\!\!\!\!\! \!\!\!\!\!\! \!\! (\ref{channels}),  (\ref{power}),\label{lk} 
\end{align}
\end{subequations}
 where $\alpha_2=\alpha$ and  $\alpha_1=1-\alpha$. It is obvious that $\overset{\approx}{P_1}$ cannot be solved directly through existing convex optimization software due to the non-convex constraints. Hence, we approximate those constraints with convex ones and iteratively  solve the SOO problem by updating the approximations in each iteration.
 Without loss of generality, the constraint in (\ref{con1}) can be   equivalently  written as
  \begin{equation}\label{kcons1}
  \sum_{i=1}^{K} {R}_i \geq \frac{f_1^{max}\xi_1}{(1-\alpha)}.
\end{equation}
  By introducing a new set of slack variables, the  constraint in (\ref{kcons1})   can be rewritten into the following set of constraints:
\begin{subnumcases}{(\ref{kcons1})  \Leftrightarrow  }
   \sum_{i=1}^{K}{r}_i\geq \frac{f_1^{max}\xi_1}{(1-\alpha)}, \label{part2}
   \\
    {R}_i  \geq  r_i,   \forall i \in \mathcal{K}. \label{saw}
\end{subnumcases}
Furthermore, the constraint in  (\ref{saw}) can be expressed as
\begin{equation}\label{rat1}
\log\left(1+\frac{\vert \mathbf{h}_k^H\mathbf{w}_i \vert^2}
{\sum_{j=i+1  }^{K}\vert \mathbf{h}_k^H \mathbf{w}_j \vert^2+\sigma_k^2}\right) \geq r_i,   \forall i \in \mathcal{K}, k\geq i.
\end{equation}
The non-convexity of (\ref{rat1}) can be handled by introducing new slack variables, namely  $z_i$ and $ \rho_i$,   such that the constraint in (\ref{rat1}) is written into the following two  constraints:
\begin{subequations}
\begin{align}
& 1+\frac{\vert \mathbf{h}_k^H\mathbf{w}_i \vert^2}
{\sum_{j=i+1  }^{K}\vert \mathbf{h}_k^H \mathbf{w}_j \vert^2+\sigma_k^2}  \geq z_i,\forall i \in \mathcal{K}, k\geq i, \label{parrt2}\\
&{z}_i  \geq  2^{r_i},\quad  \forall i \in \mathcal{K}. \label{saww}
\end{align}
\end{subequations}
In addition, the constraint in (\ref{parrt2}) is similarly written with a  new slack variable    $\eta_{i,k}$ as
\begin{subnumcases}{(\ref{parrt2})\\  \Leftrightarrow  }
    \vert \mathbf{h}_k^H\mathbf{w}_i \vert^2 \geq (z_i-1)\eta_{i,k}^2,     \label{parrtt2}
   \\
    {\sum_{j=i+1  }^{K}\vert \mathbf{h}_k^H \mathbf{w}_j \vert^2+\sigma_k^2} \leq \eta_{i,k}^2. \label{sawww}
\end{subnumcases}
To represent    (\ref{parrtt2})   in  a convex form,   the square root of each term in the inequality is considered.  Then, we approximate the right hand-side term of the inequality by  convex-concave approximation \cite{cc} through the first-order Taylor series approximation \cite{hanif}. Therefore, the constraint in  (\ref{parrtt2}) can be approximated by the following linear inequality constraint:
\begin{multline}\label{ta3ab}
{\Re (\mathbf{h}_k^H\mathbf{w}_i )}\geq \\ \sqrt{{( z_i^{(n-1)}-1)} }\eta_{i,k}^{(n-1)}+ \sqrt{ {({ z_i^{(n-1)}-1)}} }(\eta_{i,k}-\eta_{i,k}^{(n-1)})\\+0.5 \sqrt{\frac{ 1}{{( z_i^{(n-1)}-1)}}}( z_i- z_i^{(n-1)}),    \forall i \in \mathcal{K}, k\geq i,
\end{multline}
where $ z_i^{(n-1)}$ and $\eta_{i,k}^{(n-1)} $ represent the  approximations of $ z_i $ and $\eta_{i,k} $, at the $ (n-1)$ th iteration, respectively.
Furthermore, we can rewrite the constraint in (\ref{sawww}) as the following
 second order cone (SOC):
\begin{equation}\label{soc1}
\eta_{i,k} \geq ||[\mathbf{h}_k^H \mathbf{w}_{i+1}~ \mathbf{h}_k^H \mathbf{w}_{i+2}~\cdots~\mathbf{h}_k^H\mathbf{w}_{K}~\sigma_k ]^T||_2,      \forall i \in \mathcal{K}, k\geq i.
\end{equation}
With multiple slack variables, the non-convex constraint in (\ref{con1}) is now formulated as the following convex constraints:
\begin{equation}
(\ref{con1})  \Leftrightarrow  (\ref{part2}),(\ref{saww}),(\ref{ta3ab}),(\ref{soc1}).
\end{equation}
\indent Next,  we   rewrite the constraint in (\ref{con2}) in convex form by employing  the same approximation techniques that have been already implemented to handle (\ref{con1}). Firstly, we   equivalently rewrite the constraint in  (\ref{con2}) as
\begin{equation}\label{naraa}
\frac{(\sum_{i=1}^{K} r_i)^2}{K \sum_{i=1}^{K} r_i ^2}\geq \frac{\xi_2}{ \alpha},
\end{equation}
and this constraint is approximated by introducing the following new slack variables $ \gamma,\beta$ such that
\begin{subnumcases}{(\ref{naraa})  \Leftrightarrow  }
   (\sum_{i=1}^{K} r_i)^2 \geq \gamma \beta^2, \label{ar}
   \\
    {K \sum_{i=1}^{K} r_i ^2} \leq \beta^2, \label{arrd}
\\
\gamma \geq \frac{\xi_2}{ \alpha}. \label{arr}
\end{subnumcases}
To handle the non-convexity of the constraint in  (\ref{ar}), we use the same Taylor series approximation that was employed to derive the inequality in (\ref{ta3ab}),  as follows:
\begin{multline}\label{jara}
 \sum_{i=1}^{K} r_i \geq 
 \sqrt{\gamma^{(n-1)}} \beta^{(n-1)}+\\ 0.5 \frac{1}{ \sqrt{\gamma^{(n-1)}}}\beta^{(n-1)}( \gamma-\gamma^{(n-1)})   +
 \sqrt{\gamma^{(n-1)}} (\beta-\beta  ^{(n-1)}).
\end{multline}
Similarly,  the constraint in (\ref{arrd}) can be written as the following SOC constraint:
\begin{equation}\label{con5}
\beta \geq \sqrt{K}  || [r_1~r_2~\cdots~r_K]^T||_2.
\end{equation}
Hence, the non-convex constraint in (\ref{con2})   can be now reformulated as the following convex constraints:
\begin{equation}
(\ref{con2})  \Leftrightarrow  (\ref{arr}),(\ref{jara}),(\ref{con5}).
\end{equation}
 Finally, we transform the non-convex SIC constraint in (\ref{channels}) by  replacing  each non-convex term of the inequality   with  a linear approximated term
 using the first-order Taylor series approximation, as shown below \cite{haitham}
\begin{multline}\label{anoora}
\vert \mathbf{h}_k^H\mathbf{w}_j  \vert^2 \cong  |\vert [ \Re{(\mathbf{h}_k^H \mathbf{w}_j^{(n-1)})}~ \Im{(\mathbf{h}_k^H \mathbf{w}_j^{(n-1)})}]^T||^2 +\\2 [ \Re{(\mathbf{h}_k^H \mathbf{w}_j^{(n-1)})} ~ \Im{(\mathbf{h}_k^H \mathbf{w}_j^ {(n-1)})}]\\ [(\Re{(\mathbf{h}_k^H \mathbf{w}_j)}-\Re{(\mathbf{h}_k^H \mathbf{w}_j^{(n-1)})})~\\ (\Im{(\mathbf{h}_k^H \mathbf{w}_j)}-\Im{(\mathbf{h}_k^H \mathbf{w}_j^{(n-1)})})]^T.
\end{multline}
\indent Through including these approximations,   the original optimization problem $\overset{\sim}{ P_1}$ can be reformulated as follows:
\begin{subequations}\label{MOO6}
\begin{align}
& \underset{\chi  }{\text{max}}
& &  \xi \\
& \text{s.t.}
& & (\ref{channels}),(\ref{saw}) (\ref{saww}), (\ref{ta3ab}),(\ref{soc1}),     \\
& && (\ref{cons1}),(\ref{lk}),(\ref{part2}), (\ref{con5}),(\ref{arr}),(\ref{jara}),
\end{align}
\end{subequations}
where  $\chi$  includes all the  optimization parameters  such that  $\chi\overset{\bigtriangleup}{=}\{\mathbf{w}_i,\xi,\xi_1,\xi_1,\beta,\gamma,r_i,\eta_{i,k},z_i  \}_{i=1}^{K}.$  It is obvious that solving the  optimization problem in (\ref{MOO6}) requires to initialize the parameters  $\chi^{(0)}$ and these parameters can be obtained by   choosing a feasible beamforming vectors $\{\mathbf{w}_i^{(0)}\}_{i=1}^{K}$. Furthermore, the other slack variables can be determined through substituting    $\{\mathbf{w}_i^{(0)}\}_{i=1}^{K}$ in the inequalities. The  optimization problem in (\ref{MOO6})
is  iteratively solved until the required accuracy is achieved such that $|\xi^{(n)}-\xi^{(n-1)} |$ is less than a pre-defined threshold $\varrho$.

\section{Simulation Results}\label{sec4}
In this section, we provide simulation results examining the effectiveness of the proposed sum rate-fairness  trade-off-based beamforming design. In particular, we consider a BS  equipped with four transmit antennas (i.e., $N=4$) which transmits information to five single-antenna users.   It is assumed that the users are located at  distances of 50, 4, 3, 2 and 1 meters. Furthermore, we assume that all the channels are Rayleigh  fading,   with the   path loss exponent and the noise variance of the channels are set to be two and one, respectively, whereas the available bandwidth for   transmission is assumed to be  $B_w=1$ MHz. The threshold to terminate the  iterative algorithm is chosen to be 0.001 (i.e., $\varrho=0.001$).  We define the  normalized  transmit power   (TX-SNR) in dB as   $\text{TX-SNR (dB)}=10\log_{10}\frac{P_{ava}}{\sigma^{2}}. $
\begin{figure}[htb]\label{fig:vara}
 \includegraphics[scale=0.23,center]{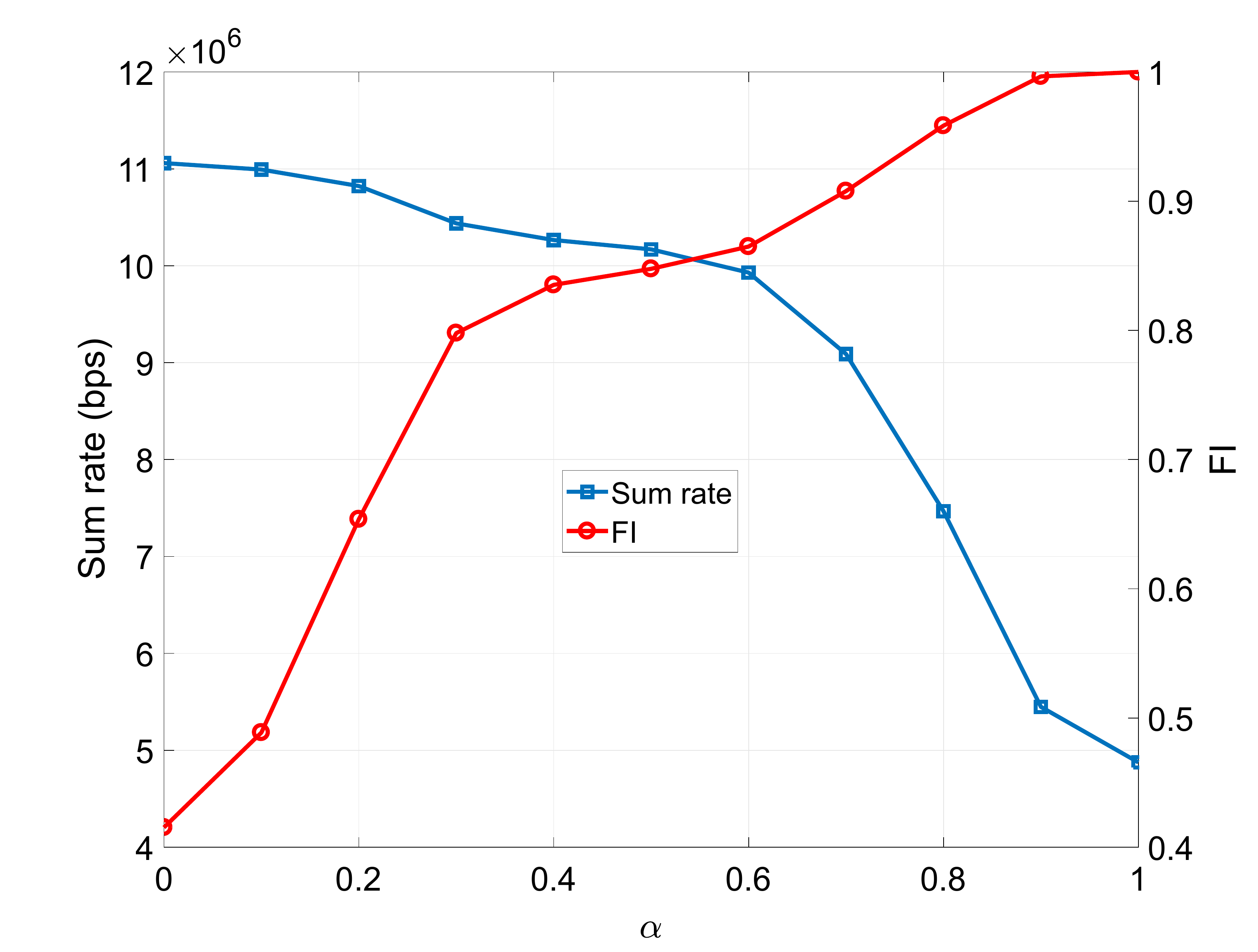}
 \caption{Achieved sum rate and FI against the weight factor $\alpha$, TX-SNR=30 dB. }
 \centering
 \label{fig:one}
 \end{figure}
Fig.  \ref{fig:one} demonstrates the achieved  sum rate and FI over the   weight factor   $\alpha$. As expected,  the problem $ \overset{\sim}{P_1}$ becomes SRM at $\alpha=0$, and   the maximum sum rate is  achieved   at the cost of    lower FI. Furthermore, the optimal fairness is achieved when the weight factor is set to be one (i.e., $\alpha=1$). In particular, the problem $ \overset{\sim}{P_1}$ turns out to be MMR with $\alpha=1$. However, the  BS  can appropriately choose a value for the weight factor $\alpha$ so that a good balance between the sum rate and FI can be achieved. A good trade-off between these performance metrics can be achieved by choosing  $\alpha=0.5$, as shown in Fig.  \ref{fig:one}.
\begin{figure}[htb!]\label{fig:vara2}
 \includegraphics[scale=0.23,center]{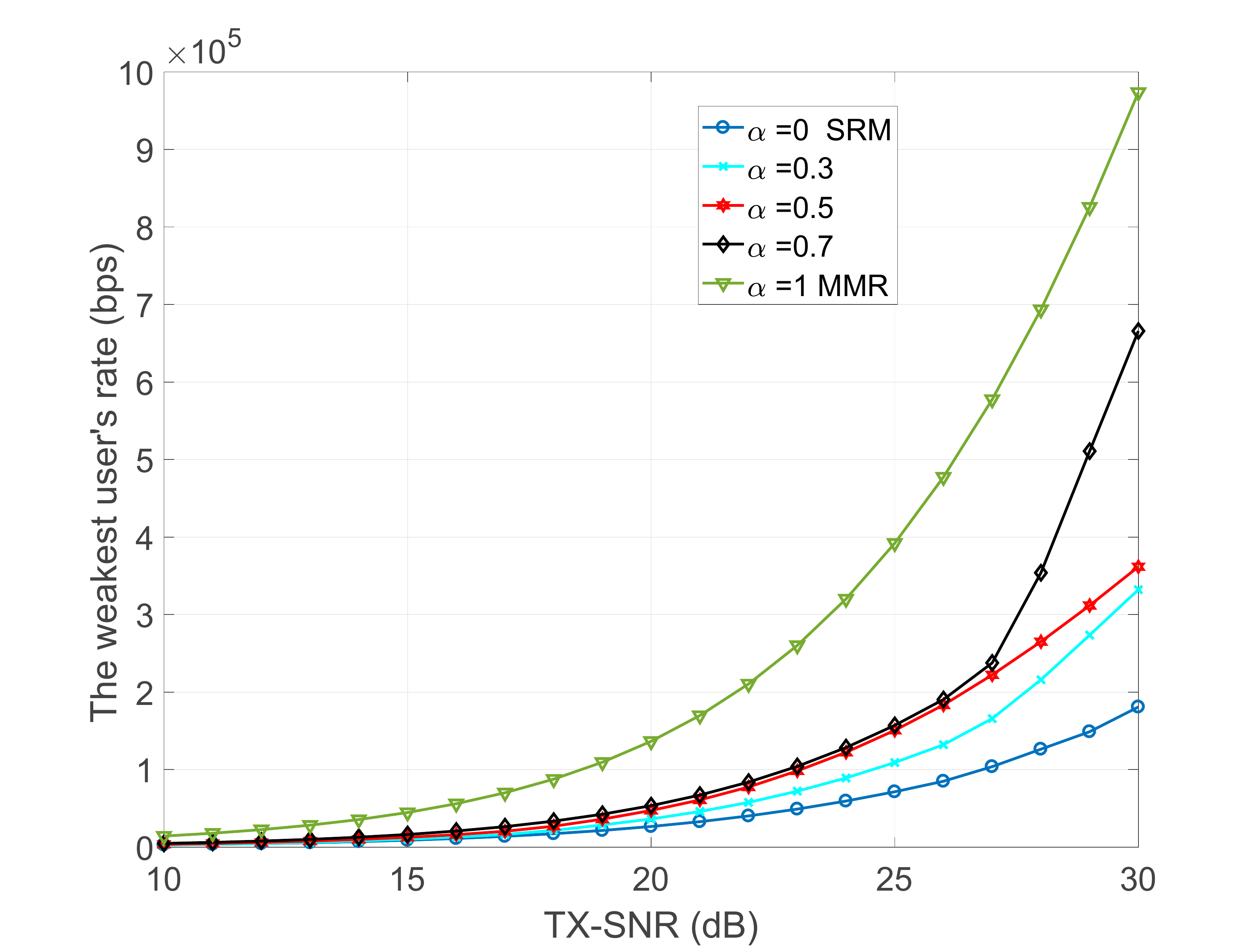}
 \caption{The weakest user's rate against the available transmit power for  different weight factors $\alpha$. }
 \centering
 \label{fig:two}
 \end{figure}
Fig.  \ref{fig:two} illustrates the rate variation of the weakest user for different values of  the weight factor $\alpha$.   For example,  at TX-SNR= 30 dB, the rate of the  weakest user  achieves around 0.2 Mbps with $\alpha=0$; however, this rate can be increased  five times by setting the weight factor $\alpha$ to   1. Hence, the BS has the flexibility to determine the  achievable rate of the weakest user by appropriately choosing the weight factor $\alpha$.
\begin{figure}[htb!]\label{fig:vara1}
 \includegraphics[scale=0.23,center]{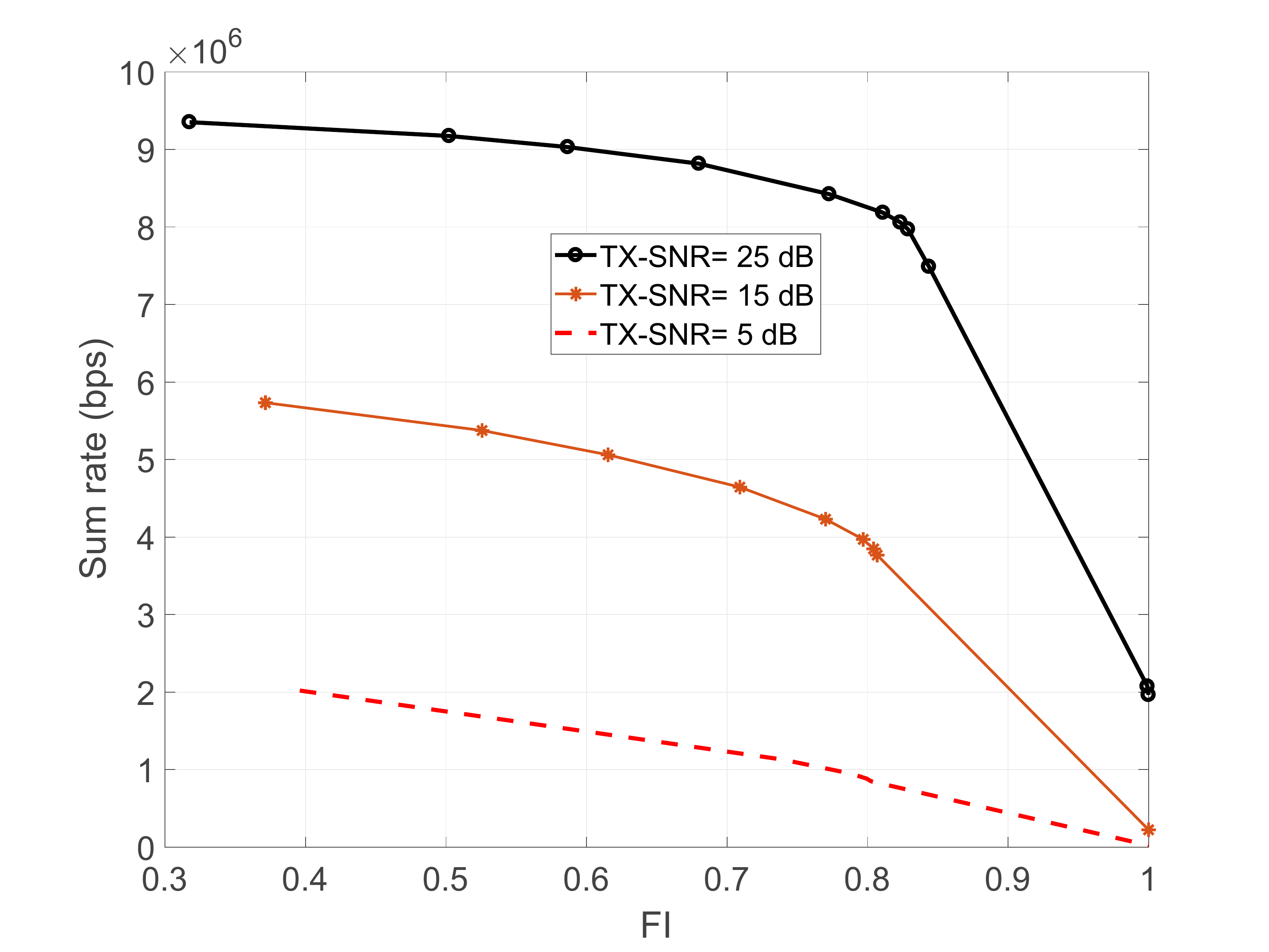}
 \caption{Pareto-front for different TX-SNR thresholds. }
 \centering
 \label{fig:three}
 \end{figure}
Furthermore, we provide the Pareto-optimal solutions of the proposed joint sum rate-fairness-based beamforming  design for different TX-SNR thresholds in Fig.  \ref{fig:three}. In particular, the Pareto-front is the set that consists of  the best-trade off  (Pareto-optimal) solutions for $ \overset{\sim}{P_1}$. For instance, at TX-SNR=25 dB, each point on the curve represents the best (sum rate, FI) solution   that could be achieved with a particular weight of $\alpha$. It is worth  mentioning that for a given value of $\alpha$, one of the  performance metrics can be improved. However, this improvement is not without  degrading the performance of the other metric.

\begin{table*}
\centering
\caption { The impact of the weakest user distance (i.e., $d_1$) on the sum rate and FI with different weight factors, at TX-SNR=35 dB.} \label{tab:title}

 \centering
  \resizebox{\textwidth}{!}{
 \begin{tabular}{|c||c|c|c|c||c|c|c|c||c| c|c| c| c|c|c|c|c|c|c|c|c||c|c||c|c||c|c|c|c|c|c|c|c|c|c||c|c|}
  \hline
    &   \multicolumn{4}{c||}{Case 1, $d_1$=10 m } &\multicolumn{4}{c||}{Case 2, $d_1$=100 m }&\multicolumn{4}{c| }{Case 3 $d_1$=1000 m }\\
     \hline
   &R (Mbps)&$R_1$(Mbps)&$R_5$ (Mbps) &FI &R (Mbps)&$R_1$(Mbps)&$R_5$ (Mbps) &FI&R (Mbps)&$R_1$(Mbps)&$R_5$ (Mbps) &FI  \\
   \hline
    $\alpha=0$ & 13.598&0.5537 &8.9750 &0.4224 &13.4003 &0.2033 &9.2729 &0.3845&13.2006 & 0.0042&9.2421 &0.3750\\
 \hline
   $\alpha=0.25$ &13.1764  &2.0968 &3.6811  & 0.9571 &12.9283 & 0.3935  & 4.2581  &  0.7974 &12.9127 &  0.0057  &  5.1300  &0.6559    \\
   \hline
    $\alpha=0.5$&12.9801 &2.3456&3.0192 &0.992&12.6967 &0.4279 &3.4141 &0.8487&12.3254 &0.0081 &3.5196 &0.7930 \\
   \hline
    $\alpha=0.75$& 12.9111  & 2.3686   &2.8195  & 0.9966&10.9714    &0.9287  & 2.5853 &0.9229 &12.2129  &  0.0082   & 3.1948   & 0.8004   \\

   \hline
    $\alpha=1$& 12.7126&2.6919 &2.4219 &0.999& 7.6435&1.1216 &1.3380 &0.9484&0.1469 &0.0340 & 0.0374&0.9951 \\
    \hline
  \end{tabular}}

\end{table*}
Finally,    Table \ref{tab:title}   demonstrates the importance of the proposed sum rate-fairness trade-off-based design over the other  fixed beamforming design. In particular, the BS decides the weights of each of the conflicting performance metrics (i.e., sum rate and FI) based on the instantaneous channel state information of the users. To explain this in a detailed manner, we present  the  rates  of the weakest and  strongest users, the achieved sum rate, and the FI through increasing the distance of  the  weakest user $u_1$ ($d_1$)  from 10-1000 meters, while  the distances of the remaining  four users in the system remain fixed.
As seen in  Table \ref{tab:title}, the SRM beamforming design (i.e., $\alpha=0$) does not provide a better quality of service  in terms of fairness and achievable rate of the weakest user as the distance between users increases. For example, the weakest user achieves only a rate of  0.0042 Mbps while the strongest user enjoys a rate of  9.2421 Mbps, with   FI of 0.375. Hence, the BS can intelligently assign appropriate weights to maintain a good fairness among the users,   such that the weakest user rate is reasonably increased.
\section{Conclusions}\label{sec5}

In this paper, we have proposed a  sum rate-fairness trade-off-based beamforming design  for a MISO NOMA system. In this  design, the BS has the flexibility to appropriately choose the weights of each objective  according to the users' channel conditions. The beamforming   design is  formulated as a MOO problem which is hard to solve directly. To overcome this issue, a weighted sum approach combined with the prior articulation method is employed to reformulate  the original  problem as a SOO problem. Furthermore,  the SCA  technique is exploited to iteratively solve the weighted SOO problem. Simulation results indicate that the proposed approach is very effective when compared against  the conventional rate-aware beamforming design.
 
{  \section*{Acknowledgement}
 The work of K. Cumanan, A.  Burr and Z. Ding was supported by H2020-MSCA-RISE-2015 under grant no: 690750. }
\bibliographystyle{IEEEtran}
\bibliography{references}

\end{document}